# 超结构化四边形网格对有限元计算收敛性和精度影响的探索

赵 辉[*]

[*] (可计算离散整体几何结构实验室，graphicsresearch@qq.com，北京)

**摘要**：在当前工业界和学术界的实践中，有限元计算的收敛性、精度和网格划分的方式和质量紧密相关。多年来，国际国内学术领域研究的高质量的网格划分主要指的是四边形、六面体的**局部的**质量近似于正方形和立方体。本文的主要贡献是提出一个全新的研究方向和研究内容：需要探索和研究超结构化四边形网格的四边形**整体全局**排列结构和方式对于有限元计算收敛性和计算精度的影响。通过这个全新领域的研究，可以有助于解决当前工业界和学术界在做仿真模拟时对于网格划分阶段严重依赖于"经验"的不严谨状态，对于哪种网格划分的全局排列可以保证有限元计算的收敛做出明确的判断。而为了能生成和设计整体排列结构可控的超结构化四边形网格（如图 1 所示），需要大量的前沿二维和三维几何拓扑理论，如模空间、泰希米勒空间、调和叶状结构、动力系统、曲面映射、亚纯二次微分、曲面映射等。

**关键词**：超结构化四边形网格，有限元计算，收敛性，计算精度，整体几何结构，调和叶状结构，泰希米勒空间

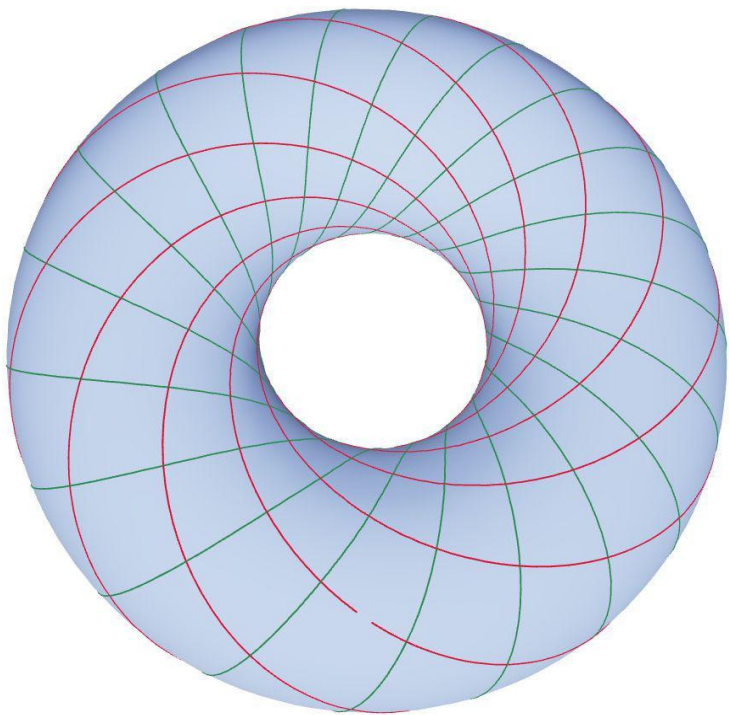

(图 1：超结构化四边形网格，赵辉用原创几何拓扑代码平台 Geometric 作图。)

(Fig.1：Super Structured Quad Mesh, Generated By Hui Zhao with the Software Geometric .)

# 引 言

模型曲面上的网格划分可以三角形、四边形、混合网格等，模型内部也分别对应四面体、六面体、混合网格。对后续的有限元计算来说，学术界工业界认为如果网格表面全部是四边形和内部的全部是六面体构成，那么相对于表面的三角形和内部的四面体，在计算收敛和精度上具有优势。但是对于同一个三维模型，可以有无数个不同的纯四边形表面和纯六面体实体网格划分。当前一个主流的观点是四边形和六面体的局部几何性质对计算有影响，例如四边形的形状，最好是逼近于完美的正方形和立方体。本文提出了一

个全新的科学问题："在局部四边形和六面体几何质量一样的情况下，仅仅因为四边形和六面体**整体排列结构**的不同造成的不同的网格划分结果是否对后续有限元计算的收敛是否有本质的影响，以及对计算精度有什么影响？如图 1 这种绕圈的四边形和对应的内部六面体整体排列结构"。对于这个科学问题，我们仍旧在研究中，我们预期研究结果将会支持这个问题答案是肯定的。

# 1 超结构化四边形网格

## 1.1 概念和定义

在学术界和工业界，网格划分通常分为非结构化（如图 2 图 3 左图）、结构化（如图 2 图 3 右图）、混合网格。非结构化指的是表面是三角形内部是四面体的网格，结构化指的是表面全部是四边形和内部全部是六面体的网格，混合网格指的是网格元素有三角形、四边形、五边形、四面体、六面体等各种形式混合而成。

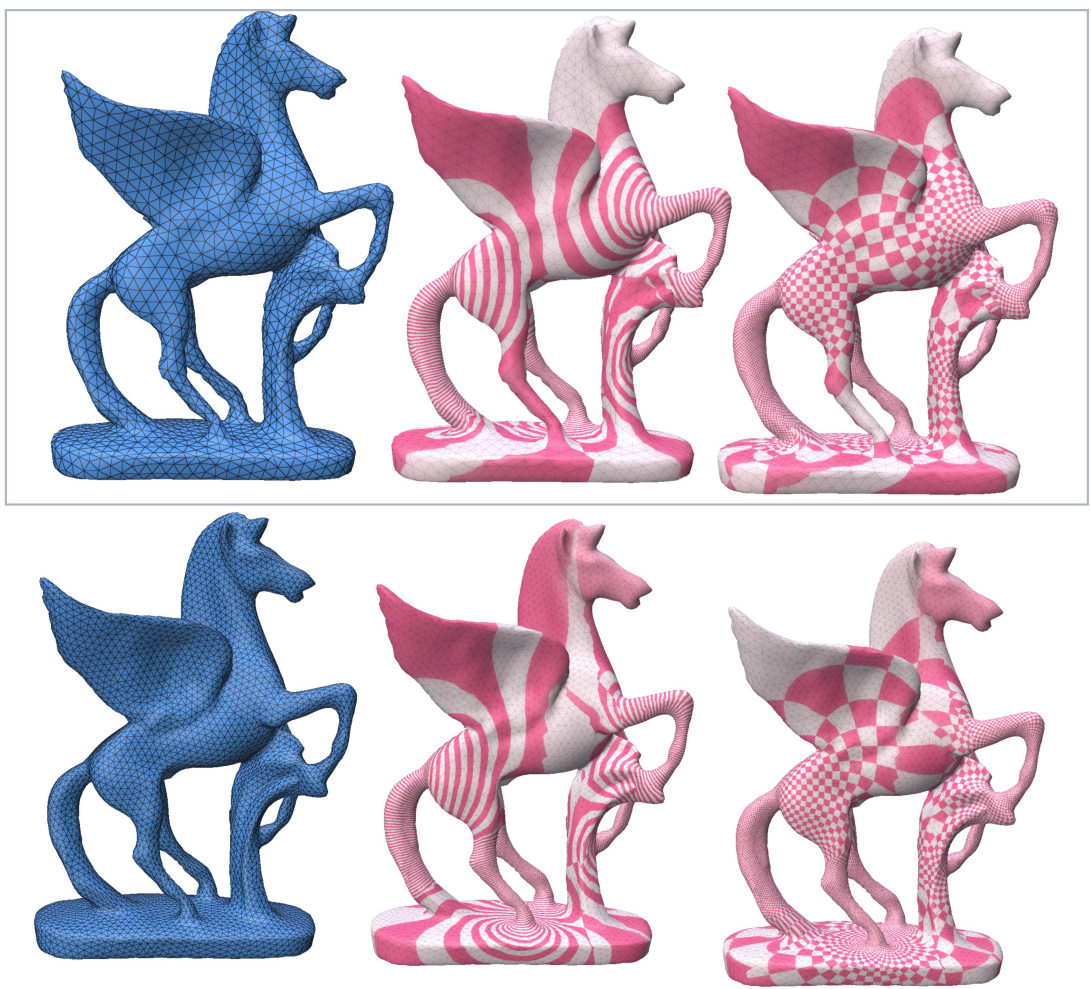

(图 2：左图三角形非结构化网格，右图结构化四边形网格，赵辉用原创几何拓扑代码平台 Geometric 作图。)

(Fig.2: left is triangle mesh, right is structured quad mesh, Generated By Hui Zhao with the Software Geometric .)

在文献[1] 对于结构化的网格划分，又进一步细分为规则化（Regular）、半规则化（Semi-Regular）、度数半规则化（Valence Semi-Regular）、非规则化（Irregular），如文献[1] 中的图所示。但这几种规则化的划分没有明确严谨的判断标准，是根据得到的区域 (blocks) 数量大致进行定义，一般来说得到的区域 (blocks)

越少越规则。也就是没有办法通过算法来对于给定的结构化网格来判断是否规则化还是非规则化，通常期望得到的区域数量越少越好。

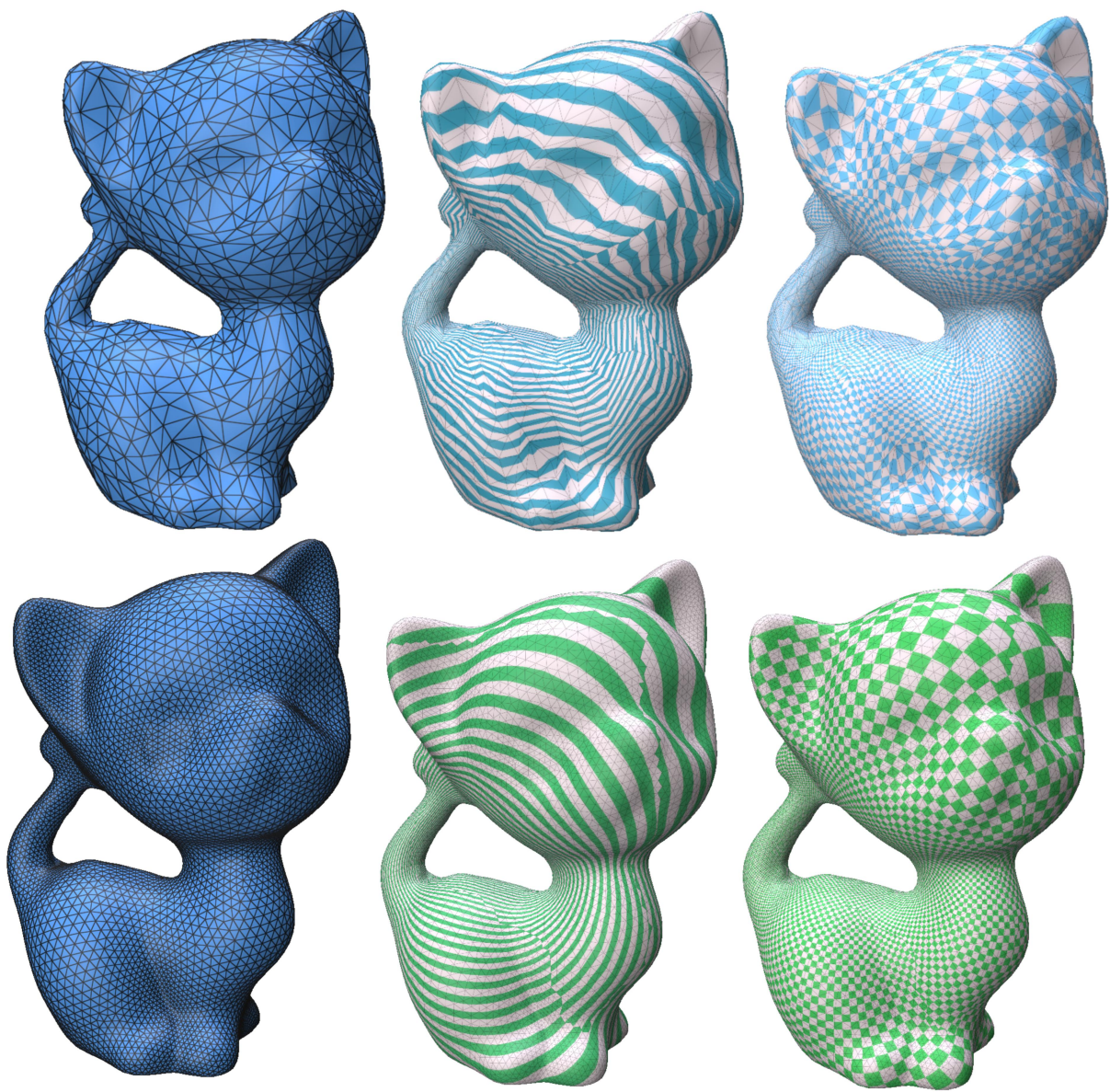

(图 3：左图三角形非结构化网格，右图结构化四边形网格，赵辉用原创几何拓扑代码平台 Geometric 作图。)

(Fig. 3：left is triangle mesh，right is structured quad mesh，Generated By Hui Zhao with the Software Geometric .)

在通常的结构化四边形网格（Structured Quad Mesh），规则化四边形网格分类的基础上，在 2023 年，我们提出了"四边形整体排列结构可控"的**超结构化四边形网格**（Super Structured Quad Mesh）的全新的概念、研究方向、研究领域、研究内容。

同一个三维模型可以对应无数种网格划分，如图 2 第一行和第二行的网格划分所示。即使对于结构化的网格来说，同一个模型，也具有无数种的结构化划分结果，每一种结果里四边形的整体排列结构都不一样，有可能是一般常见的排列方式，如图 4 所示，也有可能是绕圈的排列方式，如图 1 所示。而之前的研究中，所有的算法都仅仅是随机得到某一种划分结果，而无法精确的控制和预先设计得到哪一种排列结构。我们提出的超结构化四边形网格研究领域就是研究如何设计和控制得到的特定的整体排列结构。超结构化

四边形网格概念里的"超"就指的是对于四边形排列结构的可控可设计。

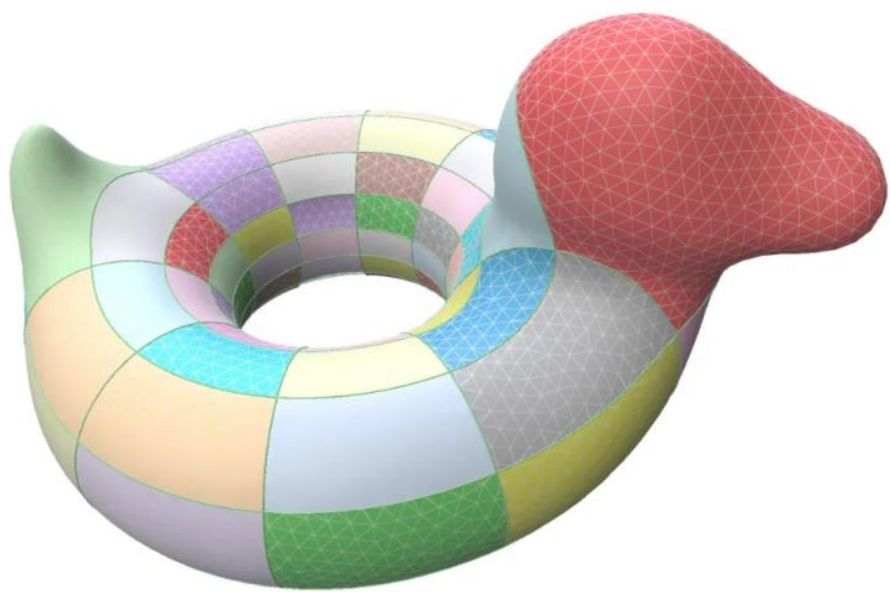

(图 4：超结构化四边形网格，赵辉用原创几何拓扑代码平台 Geometric 作图。)

(Fig. 4：Super Structured Quad Mesh，Generated By Hui Zhao with the Software Geometric .)

**1.2 当代微分几何拓扑概念和理论**

非结构化网格划分方面的研究国际国内已经有数十年历史，有大量的经典算法。四边形结构化网格划分近年来也有很多高效的算法, 这些算法基于向量场、标架场、全局无缝参数化、里奇流等理论和原理[7-30]。从非结构化到结构化的网格生成研究过程中，学术界逐渐地用到了从高中几何到古典微分几何到当代微分几何的越来越多的几何拓扑概念和理论。

而**超**结构化四边形网格的概念之所以最近几年才提出来，一个主要的原因是可控可设计整体四边形排列结构的网格需要用到更多的数学家近二十年来正在研究的新的几何拓扑理论，例如：如模空间、泰希米勒空间、动力系统、Ribbon-Graph（如图 4 所示）、调和叶状结构[2]（如图 5 所示）、全纯二次微分、亚纯二次微分、Thurston' norm、平移曲面 (Translate surface)、半平移曲面 (Half Translate surface)、平直曲面(Flat surface)、黎曼面(Riemann surface)、曲面的方块平铺(Square-tiled surfaces)、Masur-Veech volumes、Meanders（曲流形）、台球 （Billiards）、Interval exchange、Teichmuller flow、Strata of abelian and quadratic differentials、Thurston 学派的在曲面和三流形上的研究等理论。

上述这些几何结构都是整体几何结构，也就是和拓扑有关，是在曲面上全局进行定义的，相对的是一些局部定义的几何结构，如高斯曲率、平均曲率等。

表面看整体四边形排列结构可控可设计仅仅是在结构化四边形网格随机生成的结果上的一个提升，但是背后需要更多的全新几何拓扑理论进行支撑来进行算法设计。需要对上述的这些几何拓扑理论进行离散化算法设计，并能鲁棒的计算出来对应的具体数值。四边形网格可以认为是一组水平的调和叶状结构 和一组垂直的调和叶状结构两个元素构成，因此需要研究调和叶状结构的算法。在调和叶状结构算法研究过程中，涉及 Ribbon-Graph 等结构的计算。因此为了对某一个几何结构进行计算，就需要对相关的其他结构进行研究，从而上述的数十个几何拓扑概念和理论就互相联系起来。在超结构化四边形网格这个大的研究方向和研究领域里，我们正在依次地对每一个几何拓扑概念进行算法设计。例如调和叶状结构的算法，从输入的非调和叶状结构（图 7 左上），通过有约束的优化得到 Ribbon-Graph（图 5），然后再进行优化得到调和的结果（图 7 右下）。而对于全纯二次微分，可以通过在对偶模型上对输入的水平调和叶状结构旋转 90 度得到，如图 3 右下所示。

对最新发展出来的前沿几何拓扑理论在网格上进行应用可以有助于解决有限元计算、网格划分等当前

工业软件里的核心技术难题。

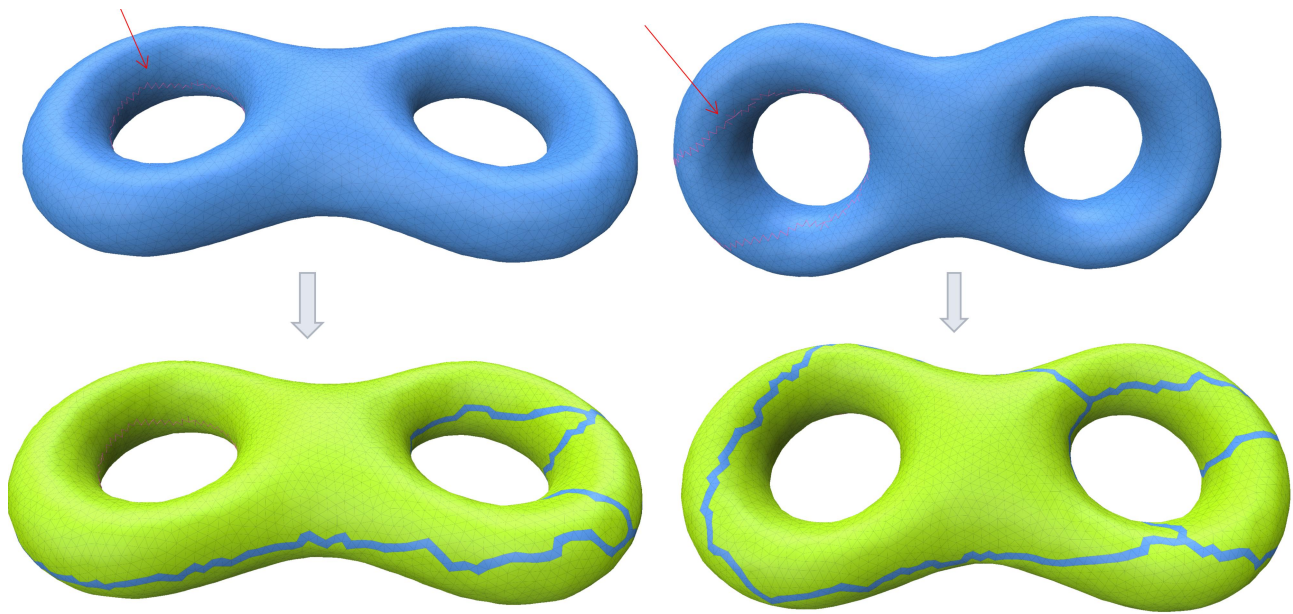

(图 5：Ribbon-Graph，赵辉用原创几何拓扑代码平台 Geometric 作图。)
(Fig. 5：Ribbon-Graph，Generated by Hui Zhao with the Software Geometric.)

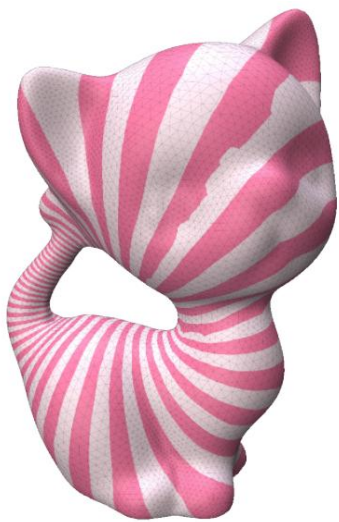

(图 6：调和叶状结构，赵辉用原创几何拓扑代码平台 Geometric 作图。)
(Fig. 6：Harmonic Foliation，Generated by Hui Zhao with the Software Geometric.)

**1.3 整体几何结构的可视化**

　　超结构化四边形网格的研究目的是工程技术角度，为了生成网格需要对各种整体几何结构进行算法设计，也就是在光滑的曲面上定义的几何结构需要在离散的网格上进行计算。这些算法设计有一个特殊的地方：离散网格上算的结果需要保持光滑曲面上定义的整体的约束或者是特征。这是当前离散网格上各种整体几何结构算法设计的难点，因为计算往往具有数值误差，很难保持整体的结构不变，所以对应的算法设计时候要以整体属性为优先考量。

　　但是能保持整体约束的算法往往很难得到，因此为了逐步对整体几何结构进行研究，可以先对整体几何结构用计算机图形学的渲染技术进行可视化，通过视觉展示出来该结构的大概规律，进行观察，然后再进一步思考和研究完善的算法。可视化的算法往往不保持整体的约束，但是在视觉上能大致体现一些规律，如图 2 图 3 右边的全纯二次微分的图示。虽然水平的调和叶状结构算法得到的结果是保持调和叶状结构的

整体属性的，但是对应的互相垂直的两个叶状结构组成的全纯二次微分当前的算法还没有能够保持整体约束，但是通过视觉的观察，可以为进一步的算法设计奠定基础。例如文献[3]里的调和微分一形式是保持整体结构的，而对应的全纯一次微分在高亏格网格上是视觉可视化的。

因此，我们在 2023 年提出可视化作为研究整体几何结构的必要条件和必要工具，而不是研究结果的附带产品。如本文中的各种图示，展现了可视化技术对于整体几何结构，以及超结构化四边形网格研究的必要性。

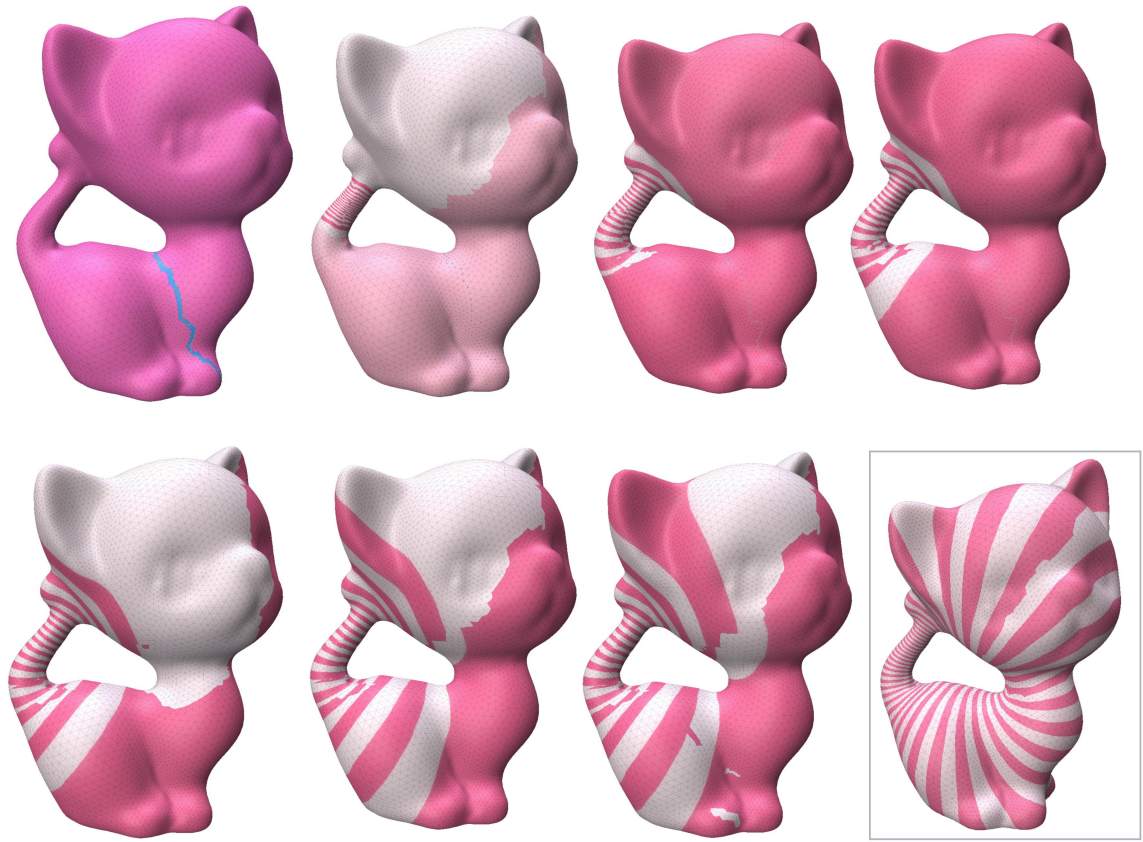

（图 7：非调和的叶状结构逐渐变为调和叶状结构，赵辉用原创几何拓扑代码平台 Geometric 作图。）
(Fig. 7: Non Harmonic to Harmonic Foliation, Generated by Hui Zhao with the Software Geometric.)

## 2 有限元计算和超结构化四边形网格

### 2.1 网格局部质量和有限元计算

网格的质量会影响有限元计算的收敛性和计算精度，一般来说，网格的元素，例如表面的三角形、四边形、内部的四面体、六面体局部的形状越规则，那么计算精度越高，收敛越快。因此当前很多工业界的网格生成软件都在优化局部形状质量的算法上下功夫。

通常来说，业界专家认为结构化的四边形和六面体网格（如图 8 左图所示）对于后续的有限元计算效率、速度、精度、收敛等方面优于非结构化三角形四面体网格和混合网格（如图 8 右图所示）。因此为了提高计算效率，往往采用结构化的四边形六面体网格。通常来说非结构化的网格比结构化的网格用到更少的几何拓扑理论，当前有很多更鲁棒和自动化的算法生成非结构化的网格。

在文献[4]里面，对结构分析(structural analysis)、热分析(thermal analysis)与低雷诺数流动(low Reynolds number flows) 等应用中的一些代表性椭圆偏微分方程在大量的网格模型上进行有限元计算对比实验，得到的结论是结构化的网格在性能上不一定优于非结构化的网格。

### 2.2 网格划分依赖于"经验"

当前在工业界还存在一个现象：对于结构化的网格，网格划分依赖于熟练工程师"经验"，如果经验不

够，可能划分出来的网格会造成后续计算模拟仿真的失败。这个依赖手工实践"经验"的情况阻碍了网格划分的自动化，影响了工业效率。这个现象也同时说明了，当前网格划分里面还存在一些几何拓扑理论没有澄清，手工实践能成功的网格划分方式和结果没有相应的数学理论解释，所以造成了没有办法用算法自动化的判断是否划分的网格可以确保后续计算的成功，以及用理论来指导设计相应的网格划分算法。

目前业界大部分认为是网格的**局部**质量造成了后续计算的失败，改进局部质量就可以使得计算成功。而我们提出的观点是，尤其在高亏格的网格上，后续计算的成败和四边形和六面体的整体排列结构有关。

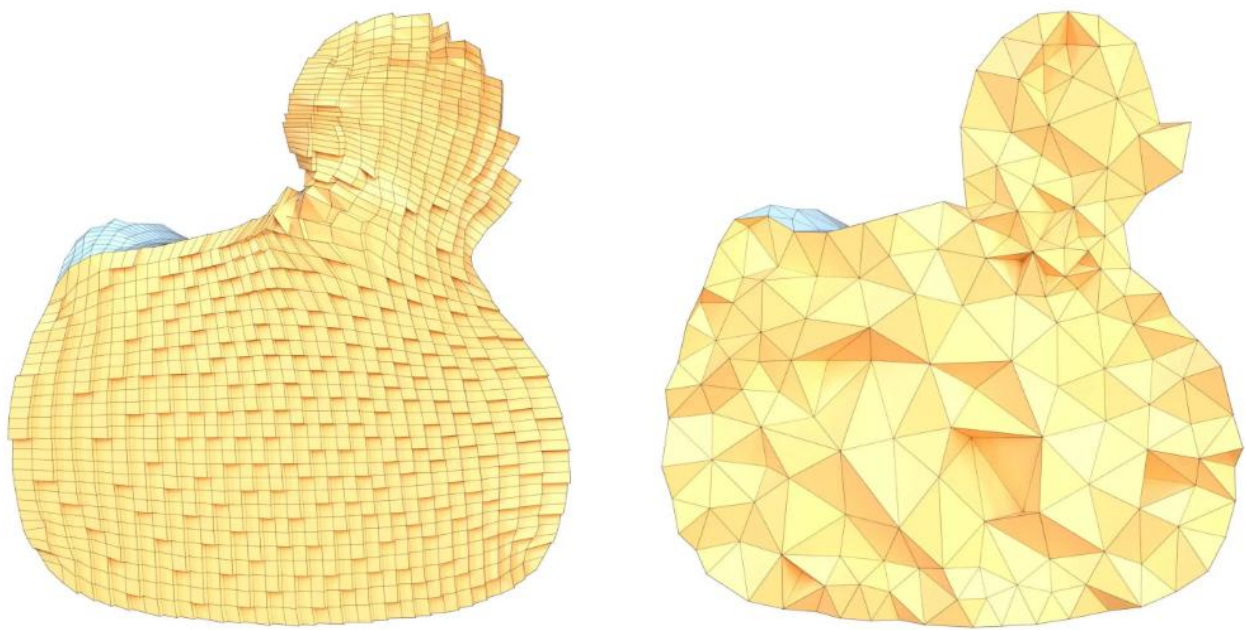

(图 8：结构化和非结构化网格，赵辉用原创几何拓扑代码平台 Geometric 作图。)
(Fig. 8: Structured vs Unstructured mesh, Generated by Hui Zhao with the Software Geometric.)

## 2.3 一个科学问题

提出一个科学问题，往往比解决一个科学问题更重要。基于在**超**结构化四边形网格上的研究，我们提出了一个全新的网格划分的整体排列结构对有限元计算影响的科学问题："**假设网格的局部几何都是规则的正方形和立方体，也就是网格局部质量都一样，对于不同四边形整体排列结构的网格和对应的六面体，是否只有某些特殊的整体排列方式才会诱导有限元计算的收敛性，并极大地提高计算精度。而其他的排列方式可能会导致有限元计算的不收敛。**"

这个科学问题的解答有助于对上述 2.1 和 2.2 里面的问题的进一步理解。当前在 CADCAE 商业软件中，需要熟练的工程师进行网格划分，才能使得后续的有限元计算收敛，如果有限元计算失败，只能重新进行划分。这种情况表明或许熟练的工程师根据经验可以对于复杂的模型划分得到我们提出的特定的"超结构化四边形网格以及与之对应的六面体网格"，只是目前工程师在实践中没有精确的几何拓扑概念来判断，只能凭经验来进行工作。

上述科学问题的解答需要对相关的前沿几何拓扑理论进行应用研究，设计鲁棒的算法生成各种亏格各种几何模型上的大量超结构化四边形网格，得到充足的实验数据，从理论到算法到实验进行详细分析，这是我们正在进行的研究工作。

## 2.4 下一代 CADCAE 工业软件的软件架构

超结构化四边形网等网格划分不仅仅和有限元计算相关，还和样条曲面设计、等几何分析、T-样条等技术密切相关。当前的商业 CADCAE 工业软件[5,6]的"软件架构"设计已经支撑软件稳定运行数十年，但

是当前仍旧存在一些技术上的难题在现有软件架构下难以解决。而这些难题的解决需要应用当代的微分几何拓扑理论，从而可能和现有的软件架构不匹配，因此我们提出以超结构化四边形网格划分为核心进行全新的下一代件架构设计，从而可以使得新一代的 CADCAE 工业软件更高效。

## 3 结论

基于对当代前沿微分几何拓扑理论的应用和算法设计，本文提出了四边形整体排列结构可控的超结构化四边形网格的概念，研究方向、研究领域，展示了初步的网格划分研究成果图示，并进一步提出了超结构化四边形网格对于有限元计算影响的科学问题。这个科学问题的研究有助于促进当前 CADCAE 等工业软件的面临的核心技术难题，促进它们的进一步发展。 更多的超结构化四边形网格划分的实验数据（如图 9 所示）可以通过电子邮箱获取[31]。

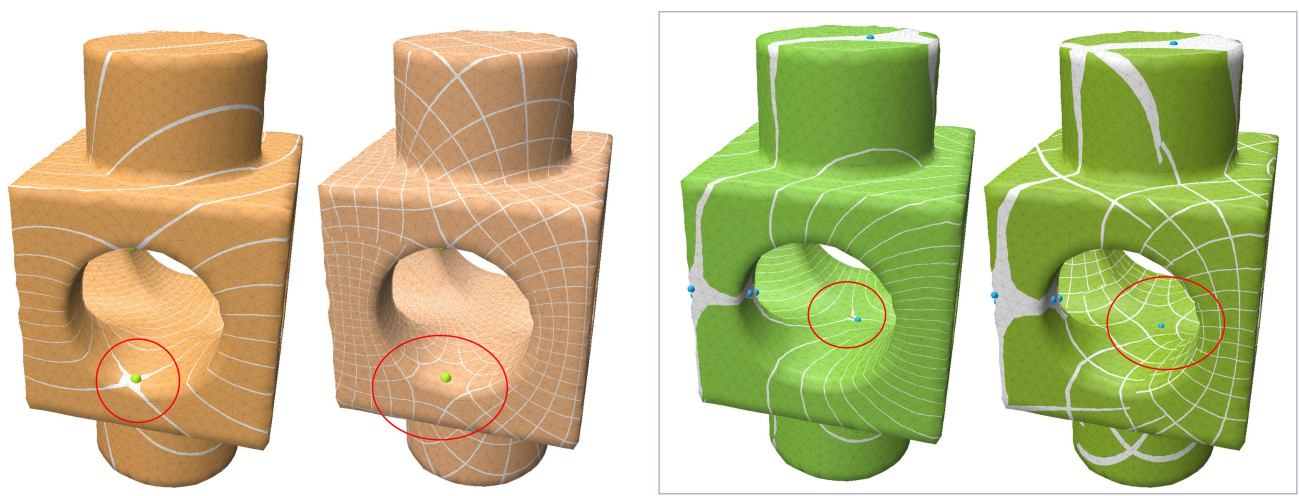

(图 9：结构化和非结构化网格，赵辉用原创几何拓扑代码平台 Geometric 作图。)
(Fig. 9: Structured vs Unstructured mesh, Generated by Hui Zhao with the Software Geometric.)

# 参考文献

1  Nico Pietroni, Marcel Campen, Alla Sheffer, Gianmarco Cherchi, David Bommes, Xifeng Gao, Riccardo Scateni, Franck Ledoux, Jean Remacle, and Marco Livesu. 2023. Hex-Mesh Generation and Processing: A Survey. ACM Trans. Graph. 42, 2 (April 2023), 1 – 44.

2  Hui Zhao, Shaodong Wang, and Wencheng Wang. 2022. Global Conformal Parameterization via an Implementation of Holomorphic Quadratic Differentials. IEEE Trans Vis Comput Graph 28, 3 (March 2022), 1529 – 1544.

3  Xianfeng Gu and Shing-Tung Yau. Global conformal surface parameterization．Proceedings of the 2003 Eurographics/ACM SIGGRAPH symposium on Geometry processing.

4  Teseo Schneider, Yixin Hu, Xifeng Gao, Jérémie Dumas, Denis Zorin, and Daniele Panozzo. 2022. A Large-Scale Comparison of Tetrahedral and Hexahedral Elements for Solving Elliptic PDEs with the Finite Element Method. ACM Trans. Graph. 41, 3 (June 2022), 1 – 14.

5  Altair.2022.HyperMesh. https://www.altair.com/hypermesh/

6  ANSYS.2022.ANSYS. https://www.ansys.com/products/meshing

7  Ryan Capouellez and Denis Zorin. 2024. Seamless Parametrization in Penner Coordinates. ACM Trans. Graph. 43, 4 (July 2024), 1 – 13.

8  David Bommes, Marcel Campen, Hans-Christian Ebke, Pierre Alliez, and Leif Kobbelt. 2013a. Integer-grid maps for reliable quad meshing. ACM Transactions on Graphics　(TOG) 32, 4 (2013), 1 – 12.

9  David Bommes, Bruno Lévy, Nico Pietroni, Enrico Puppo, Claudio Silva, Marco Tarini, and Denis Zorin. 2013b. Quad-mesh generation and processing: A survey. In Computer graphics forum, Vol. 32. Wiley Online Library, 51 – 76.

10  David Bommes,HenrikZimmer,andLeifKobbelt.2009. Mixed-integer quadrangulation. ACMtransactions on graphics (TOG) 28, 3 (2009), 1 – 10.

11  Alon Bright, Edward Chien, and Ofir Weber. 2017. Harmonic Global Parametrization with Rational Holonomy.ACM Trans.Graph.36, 4 (2017).

12  Marcel Campen. 2017. Partitioning surfaces into quadrilateral patches: A survey. In Computer graphics forum, Vol. 36. Wiley Online Library, 567 – 588.

13  Marcel Campen, David Bommes, and Leif Kobbelt. 2015. Quantized global parametrization. Acm Transactions On Graphics (tog) 34, 6 (2015), 1 – 12.

14  Marcel Campen, Ryan Capouellez, Hanxiao Shen, Leyi Zhu, Daniele Panozzo, and Denis Zorin. 2021. Efficient and Robust Discrete Conformal Equivalence with Boundary. ACM Trans. Graph. 40, 6, Article 261 (dec 2021), 16 pages.

15  Marcel Campen and Denis Zorin. 2017. Similarity maps and field-guided T-splines: a perfect couple. ACM Transactions on Graphics (TOG) 36, 4 (2017), 1 – 16.

16  Ryan Capouellez and Denis Zorin. 2023. Metric Optimization in Penner Coordinates. ACMTrans. Graph. 42, 6, Article 234 (dec 2023), 19 pages.

17  Ryan Capouellez and Denis Zorin. 2024. Seamless Parametrization in Penner Coordinates. ACM Trans. Graph. 43, 4, Article 61 (jul 2024), 13 pages.

18  Hans-Christian Ebke, David Bommes, Marcel Campen, and Leif Kobbelt. 2013. QEx: robust quad mesh extraction. ACM Trans. Graph. 32, 6, Article 168 (Nov. 2013), 10 pages.

19  Mark Gillespie, Boris Springborn, and Keenan Crane. 2021. Discrete conformal equivalence of polyhedral surfaces. ACM Transactions on Graphics (TOG) 40, 4 (2021), 1 – 20.

20  Xianfeng Gu, Ren Guo, Feng Luo, Jian Sun, and Tianqi Wu. 2018a. A discrete uniformization theorem for polyhedral surfaces II. Journal of Differential Geometry 109, 3 (2018), 431 – 466.

21  Xianfeng Gu, Feng Luo, Jian Sun, and Tianqi Wu. 2018b. A discrete uniformization theorem for polyhedral surfaces. Journal of Differential Geometry 109, 2 (2018), 223 – 256.

22  Yixin Hu, Teseo Schneider, Bolun Wang, Denis Zorin, and Daniele Panozzo. 2020. Fast Tetrahedral Meshing in the Wild. ACM Trans. Graph. 39, 4, Article 117 (July 2020), 18 pages.

23  Yixin Hu, Qingnan Zhou, Xifeng Gao, Alec Jacobson, Denis Zorin, and Daniele Panozzo. 2018. Tetrahedral meshing in the wild. ACM Trans. Graph. 37, 4 (2018), 60 – 1.

24  Jingwei Huang, Yichao Zhou, Matthias Niessner, Jonathan Richard Shewchuk, and Leonidas J Guibas. 2018. Quadriflow: A scalable and robust method for quadrangulation. In Computer Graphics Forum, Vol. 37. Wiley Online Library, 147 – 160.

25  Liliya Kharevych, Boris Springborn, and Peter Schröder. 2006. Discrete conformal mappings via circle patterns. ACM Trans. Graph. 25 (April 2006), 412 – 438. Issue 2.

26  Zohar Levi. 2022. Seamless parametrization of spheres with controlled singularities. In Computer Graphics Forum, Vol. 41. Wiley Online Library, 57 – 68.

 Zohar Levi. 2023. Seamless Parametrization with Cone and Partial Loop Control. ACM Transactions on Graphics 42, 5 (2023), 1 – 22.

27  Hui Zhao and Steven J. Gortler. 2016. A Report on Shape Deformation with a Stretching and Bending Energy. arXiv:1603.06821 [cs] (March 2016).

28  Hui Zhao, Kehua Su, Chenchen Li, Boyu Zhang, Lei Yang, Na Lei, Xiaoling Wang, Steven J. Gortler, and Xianfeng Gu. 2020. Mesh Parametrization Driven by Unit Normal Flow. Computer Graphics Forum 39, 1 (February 2020), 34 – 49. https://doi.org/10.1111/cgf.13660

29  Hui Zhao, Xuan Li, Wencheng Wang, Xiaoling Wang, Shaodong Wang, Na Lei, and Xiangfeng Gu. 2019. Polycube Shape Space. Computer Graphics Forum 38, 7 (October 2019), 311 – 322. https://doi.org/10.1111/cgf.13839

30  Hui Zhao, Na Lei, Xuan Li, Peng Zeng, Ke Xu, and Xianfeng Gu. 2017. Robust Edge-Preserved Surface Mesh Polycube Deformation. Pacific Graphics Short Papers (2017), 6 pages. https://doi.org/10.2312/PG.20171319

31  赵辉. 微信订阅号：可计算离散整体几何结构. https://mp.weixin.qq.com/s/bydA_SMV-Hu1VHYOhrvsGA?scene=1.

# A Scientist Question: Research on the Impact of Super Structured Quadrilateral Meshes on Convergence and Accuracy of Finite Element Analysis

Hui Zhao[*]

[*]( *The Lab of Computational Discrete Global Geometric Structures*,  Beijing,  China )

**Abstract**：In the current practices of both industry and academia, the convergence and accuracy of finite element calculations are closely related to the methods and quality of mesh generation. For years, the research on high-quality mesh generation in the domestic academic field has mainly referred to the local quality of quadrilaterals and hexahedrons approximating that of squares and cubes. The main contribution of this paper is to propose a brand-new research direction and content: it is necessary to explore and study the influence of the overall global arrangement structure and pattern of super structured quadrilateral meshes on the convergence and calculation accuracy of finite element calculations. Through the research in this new field, it can help solve the non-rigorous state of serious reliance on "experience" in the mesh generation stage during simulation in the current industry and academia, and make clear judgments on which global arrangements of mesh generation can ensure the convergence of finite element calculations. In order to generate and design super-structured quadrilateral meshes with controllable overall arrangement structures, a large number of modern two-dimensional and three-dimensional geometric topology theories are required, such as moduli space, Teichmüller space, harmonic foliations, dynamical systems, surface mappings, meromorphic quadratic differentials, surface mappings, etc.

**Key words**：super structured quadrilateral meshes、finite element analysis、convergence、global geometric structure、harmonic foliation、Teichmüller space

## Introduction

Mesh generation on model surfaces can be triangular, quadrilateral, or hybrid meshes, with corresponding tetrahedral, hexahedral, or hybrid meshes for the interior of the model. For subsequent finite element computations, both academic and industrial communities recognize that meshes composed entirely of quadrilateral surface elements and hexahedral volume elements offer advantages in terms of computational convergence and accuracy compared to those with triangular surface elements and tetrahedral volume elements. However, for a given three-dimensional model, there exist infinitely many distinct meshing results consisting of purely quadrilateral surfaces and purely hexahedral volumes. A prevailing view currently holds that the local geometric properties of quadrilaterals and hexahedrons influence computations; for instance, the optimal shape of a quadrilateral is one that approximates a perfect square, and similarly, a hexahedron should ideally approximate a perfect cube. This paper presents a novel scientific question: "Under the condition that the local geometric quality of quadrilaterals and hexahedrons is identical, do different overall arrangement structures of quadrilaterals and hexahedrons (which lead to distinct meshing results) have an essential impact on the convergence of subsequent finite element computations, and what influence do they exert on computational accuracy?" (An example is the looped arrangement of quadrilaterals and the corresponding overall structure of internal hexahedrons as shown in Figure 1.) We are still conducting research on this scientific question, and we anticipate that the findings will support an affirmative answer to this question.

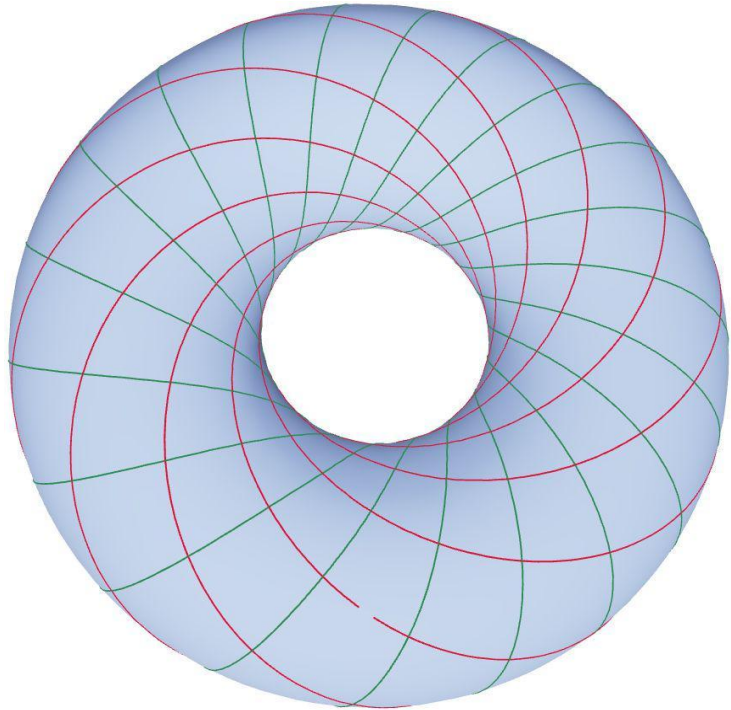

(Fig.1: Super Structured Quad Mesh, Generated By Hui Zhao with the Software Geometric .)

# 1　Super Structured Quad Mesh

## 1.1 Definition and Concept

In academia and industry, mesh generation is generally categorized into unstructured meshes (as shown in Figure 2 and the left panel of Figure 3), structured meshes (as shown in Figure 2 and the right panel of Figure 3), and hybrid meshes.

(1) Unstructured meshes refer to those where the surface consists of triangular elements and the interior is composed of tetrahedral elements.

(2) Structured meshes refer to those where the entire surface is made up of quadrilateral elements and the interior is entirely composed of hexahedral elements.

(3) Hybrid meshes refer to meshes formed by a mixture of various element types, such as triangles, quadrilaterals, pentagons, tetrahedrons, hexahedrons, etc.

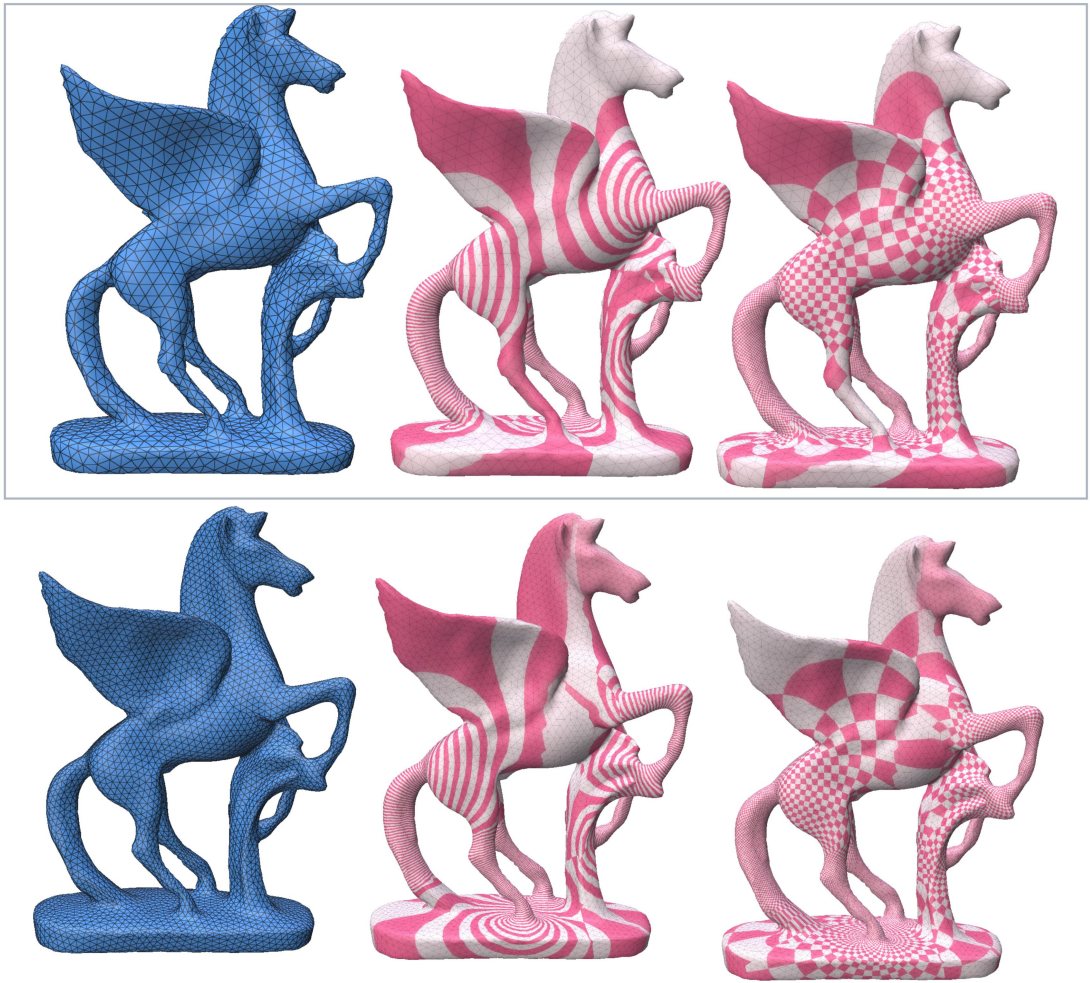

(Fig.2: left is triangle mesh, right is structured quad mesh, Generated By Hui Zhao with the Software Geometric .)

In reference[1], structured meshing is further subdivided into Regular, Semi-Regular, Valence Semi-Regular, and Irregular categories, as illustrated in the figures of reference[1]. However, there are no clear and rigorous criteria for distinguishing these types of regularization. They are roughly defined based on the number of resulting regions (blocks); generally speaking, fewer regions (blocks) indicate a higher degree of regularity. That is, there is no algorithm available to determine whether a given structured mesh is regular or irregular. It is usually desirable to minimize the number of regions.

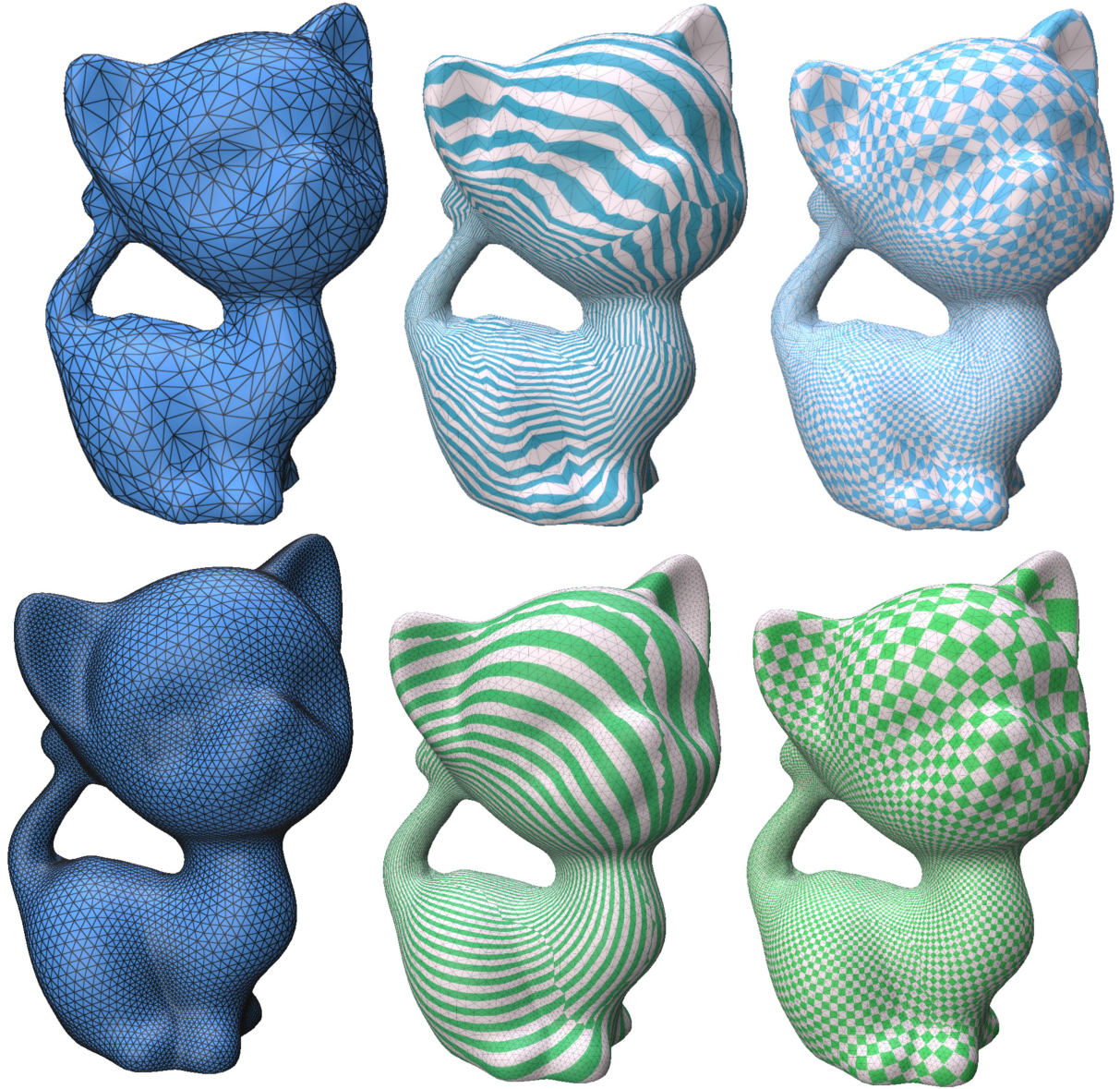

(Fig. 3:left is triangle mesh, right is structured quad mesh, Generated By Hui Zhao with the Software Geometric .)

Building on the classification of regularized quadrilateral meshes within the framework of conventional structured quad meshes, in 2023, we proposed a novel concept, research direction, research field, and research content of "Super Structured Quad Mesh" characterized by "controllable global arrangement structure of quadrilaterals". A single 3D model can correspond to countless mesh generation results, as shown in the meshes in the first and second rows of Figure 2. Even for structured meshes, a single model can have numerous structured meshing results, each with a distinct overall arrangement structure of quadrilaterals. These structures can be common arrangement patterns, such as that shown in Figure 4, or looped arrangements, such as that shown in Figure 1. However, in previous studies, all algorithms could only randomly obtain a certain meshing result, without being able to precisely control or pre-design which arrangement structure to obtain. The research field of the super structured quad mesh we proposed focuses on studying how to design and control the specific overall arrangement structure obtained. The term "super" in the concept of super structured quad mesh refers to the controllability and designability of the global arrangement structure of quadrilaterals.

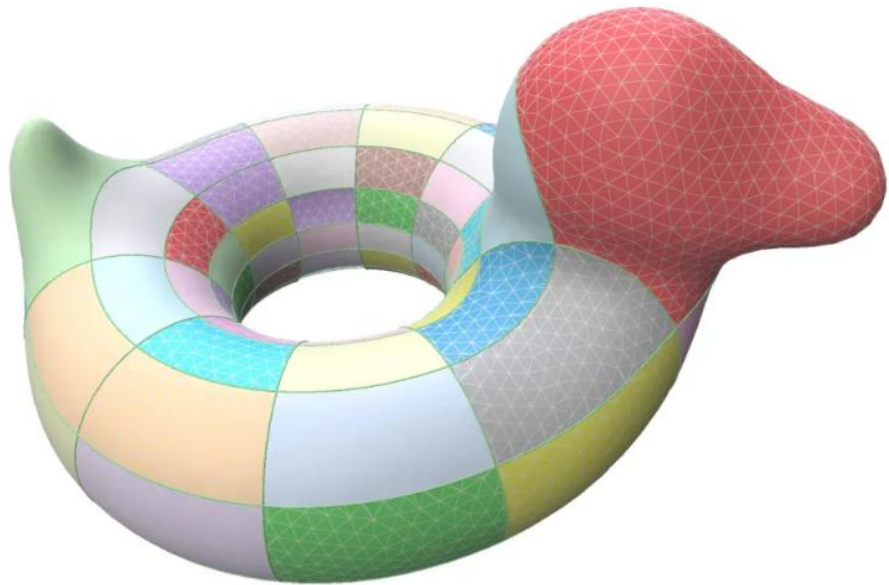

(Fig. 4: Super Structured Quad Mesh, Generated By Hui Zhao with the Software Geometric .)

**1.2 Modern Differential Geometry and Topology**

Research on unstructured mesh generation has a decades-long history both internationally and domestically, with a wealth of classical algorithms. In recent years, numerous efficient algorithms have also emerged for structured quadrilateral mesh generation, which are based on theories and principles such as vector fields, frame fields, global seamless parameterization, and Ricci flow[7-30]. In the process of research on mesh generation from unstructured to structured, the academic community has gradually adopted an increasing number of geometric and topological concepts and theories, ranging from high school geometry to classical differential geometry and contemporary differential geometry.

The concept of super structured quadrilateral meshes has only been proposed in recent years, mainly because meshes with controllable and designable overall quadrilateral arrangement structures require more new geometric and topological theories that mathematicians have been studying over the past two decades. These include, for example: moduli spaces, Teichmüller spaces, dynamical systems, Ribbon-Graphs (as shown in Figure 4), harmonic foliations[2] (as shown in Figure 5), holomorphic quadratic differentials, meromorphic quadratic differentials, Thurston's norm, translation surfaces, half-translation surfaces, flat surfaces, Riemann surfaces, square-tiled surfaces of curved surfaces, Masur-Veech volumes, meanders, billiards, interval exchanges, Teichmüller flow, strata of abelian and quadratic differentials, and the research on surfaces and 3-manifolds by the Thurston school, among others.

These geometric structures mentioned above are all global geometric structures, meaning they are related to topology and defined globally on the surface. In contrast, there are some locally defined geometric structures, such as Gaussian curvature and mean curvature.

On the surface, the controllability and designability of the global quadrilateral arrangement structure seem to be merely an improvement over the randomly generated results of structured quadrilateral meshes. However, they require more brand-new geometric and topological theories as support for algorithm design. It is necessary to design discretization algorithms for these geometric and topological theories and to robustly compute the corresponding specific values. A quadrilateral mesh can be regarded as consisting of two elements: a set of horizontal harmonic foliations and a set of vertical harmonic foliations. Therefore, it is necessary to study algorithms for harmonic foliations. In the research process of harmonic foliation algorithms, the computation of structures such as Ribbon-Graphs is involved. Thus, to compute a certain geometric structure, it is necessary to study other related structures, leading to interconnections among the dozens of geometric and topological concepts and theories mentioned above. In the broad research direction and field of super structured quadrilateral meshes, we are sequentially designing algorithms for each geometric and topological concept. For instance, for the algorithm of harmonic foliations, starting from an input non-harmonic foliation (Figure 7, top left), a Ribbon-Graph (Figure 5) is obtained through constrained optimization, and then further optimization is performed to achieve the harmonic result (Figure 7, bottom right). As for holomorphic quadratic differentials, they can be obtained by rotating the input horizontal harmonic foliation by 90 degrees on the dual model, as shown in Figure 3, bottom right.

The application of the latest frontier geometric and topological theories in meshes can help solve core technical problems in current industrial software, such as finite element computation and mesh generation.

**1.3 The Visulization of Global Geometric Structures**

The research purpose of super structured quadrilateral meshes, from an engineering and technical perspective, is to design algorithms for various global geometric structures to generate meshes. Specifically, geometric structures defined on smooth surfaces need to be computed on discrete meshes. A distinctive feature of these algorithm designs is that the results calculated on discrete meshes must preserve the global constraints or characteristics defined on the smooth surfaces. This represents a current challenge in designing algorithms for various global geometric structures on discrete meshes. Due to the inevitable numerical errors in computations, it is extremely difficult to maintain the integrity of global structures. Therefore, algorithm design must prioritize the preservation of global properties.

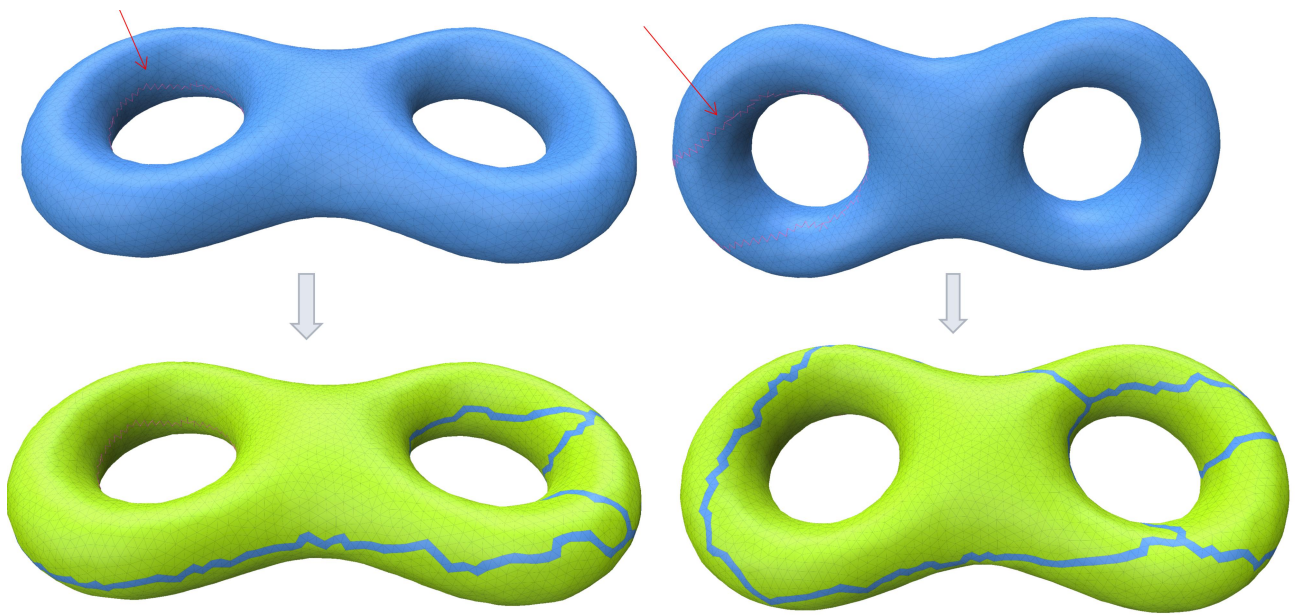

(Fig. 5: Ribbon-Graph, Generated by Hui Zhao with the Software Geometric.)

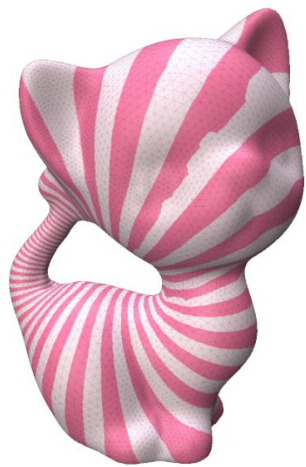

(Fig. 6: Harmonic Foliation, Generated by Hui Zhao with the Software Geometric.)

However, algorithms capable of maintaining global constraints are often difficult to develop. To gradually advance the research on global geometric structures, a feasible approach is to first visualize these structures using computer graphics rendering techniques. This allows the general patterns of the structures to be visually displayed and observed, laying the groundwork for further refinement of algorithms. Visualization algorithms typically do not preserve global constraints but can roughly illustrate certain patterns visually, as seen in the illustrations of holomorphic quadratic differentials on the right side of Figures 2 and 3. While the algorithm for horizontal harmonic foliations yields results that preserve the global properties of harmonic foliations, the current algorithm for holomorphic quadratic differentials—composed of two mutually perpendicular foliations—has not yet achieved the preservation of global constraints. Nevertheless, visual observations can provide a foundation for further algorithmic development. For example, the harmonic differential 1-forms in reference[3] preserve global structures, whereas the corresponding holomorphic 1-forms on high-genus meshes are presented through visual visualization.

Thus, in 2023, we proposed that visualization serves as an essential and necessary condition and tool for researching global geometric structures, rather than a mere accessory to research results. As demonstrated by the various illustrations in this paper, visualization technology is indispensable for the study of global geometric structures and super structured quadrilateral meshes.

## 2  Finite Element Analysis and Super Structured Quadrilateral Meshes

### 2.1 Local Quality of Mesh and Finite Element Analysis

The quality of a mesh affects the convergence and computational accuracy of finite element computations. Generally speaking, the more regular the local shape of mesh elements (such as surface triangles, quadrilaterals, and internal tetrahedrons, hexahedrons), the higher the computational accuracy and the faster the convergence. Therefore, many industrial mesh generation software tools currently focus on optimizing algorithms for local shape quality.

Typically, industry experts believe that structured quadrilateral and hexahedral meshes (as shown in the left panel of Figure 8) are superior to unstructured triangular/tetrahedral meshes and hybrid meshes (as shown in the right panel of Figure 8) in terms of subsequent finite element computation efficiency, speed, accuracy, and convergence. Thus, to improve computational efficiency, structured quadrilateral and hexahedral meshes are often adopted. In general, unstructured meshes involve fewer geometric and topological theories than structured meshes, and there are currently many more robust and automated algorithms for generating unstructured meshes.

In reference[4], comparative finite element computation experiments were conducted on a large number of mesh models for some representative elliptic partial differential equations in applications such as structural analysis, thermal analysis, and low Reynolds number flows. The conclusion drawn is that structured meshes are not necessarily superior to unstructured meshes in terms of performance.

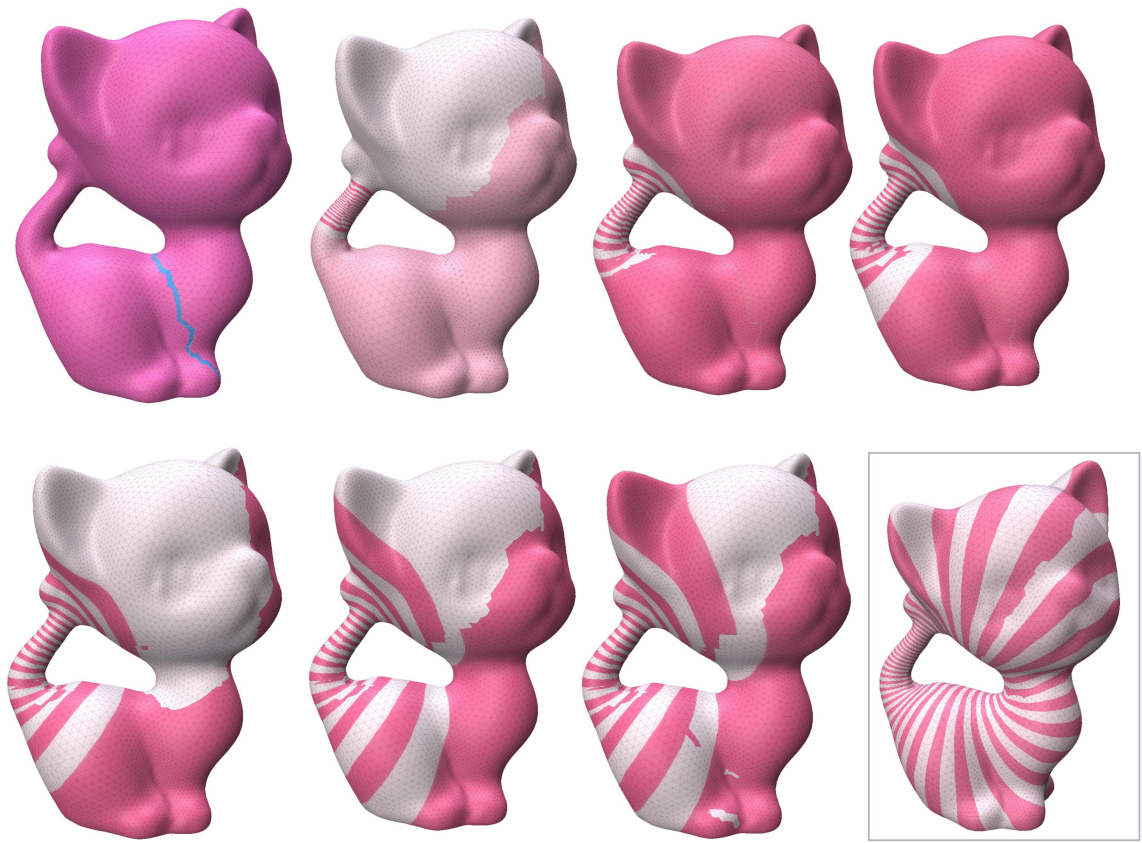

(Fig. 7: Non-Harmonic to Harmonic Foliation, Generated by Hui Zhao with the Software Geometric.)

## 2.1 Mesh generation relies on "experience"

Currently, there is a phenomenon in industry: for structured meshes, mesh generation relies on the "experience" of skilled engineers. Insufficient experience may result in meshes that lead to failures in subsequent computational simulations. This reliance on manual practical "experience" hinders the automation of mesh generation and affects industrial efficiency. This phenomenon also indicates that there are still some unclear geometric and topological theories in current mesh generation. The successful mesh generation methods and results achieved through manual practice lack corresponding mathematical theoretical explanations, making it impossible to use algorithms to automatically determine whether a generated mesh can ensure the success of subsequent computations, or to use theories to guide the design of corresponding mesh generation algorithms.

Most in the industry currently believe that the local quality of the mesh causes failures in subsequent computations, and improving local quality can ensure computational success. However, the viewpoint we propose is that, especially for high-genus meshes, the success or failure of subsequent computations is related to the overall arrangement structure of quadrilaterals and hexahedrons.

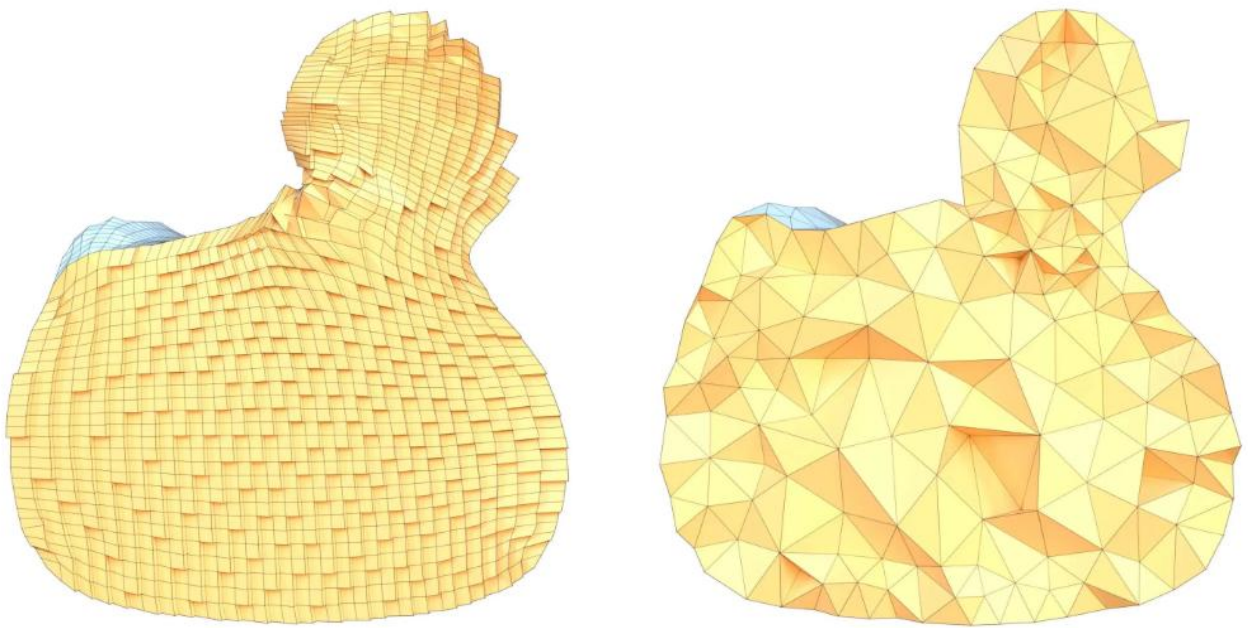

(Fig. 8: Structured vs Unstructured mesh, Generated by Hui Zhao with the Software Geometric.)

## 2.1 A Scientific Question

It is often more important to propose a scientific question than to solve one. Based on research on super structured quadrilateral meshes, we put forward a new scientific question regarding the impact of the overall arrangement structure of mesh generation on finite element computations: "Assuming that the local geometry of the mesh consists of regular squares and cubes (i.e., the local quality of the mesh is identical), for meshes with different global arrangement structures of quadrilaterals and their corresponding hexahedrons, do only certain specific global arrangements induce the convergence of finite element computations and significantly improve computational accuracy, while other global arrangements may lead to non-convergence of finite element computations?"

The answer to this scientific question will contribute to a deeper understanding of the issues discussed in Sections 2.1 and 2.2. Currently, in commercial CAD/CAE software, skilled engineers are required to perform mesh generation to ensure the convergence of subsequent finite element computations. If the finite element computation fails, the mesh must be re-generated. This situation suggests that skilled engineers may, based on experience, generate the specific "super structured quadrilateral meshes and their corresponding hexahedral meshes" we proposed for complex models. However, engineers currently lack precise geometric and topological concepts to judge such meshes in practice, relying solely on experience to carry out their work.

Answering the aforementioned scientific question requires applied research on relevant cutting-edge geometric and topological theories, designing robust algorithms to generate a large number of super structured quadrilateral meshes for geometric models of various genera, and obtaining sufficient experimental data. A detailed analysis from theory to algorithms to experiments is needed, which is the research work we are currently undertaking.

## 2.1 Software Architecture of Next Generation CAD/CAE Industrial Software

Super structured quadrilateral meshes and other types of mesh generation are not only related to finite element computation but also closely linked to technologies such as spline surface design, isogeometric analysis, and T-splines. The "software architecture" design of current commercial CAD/CAE industrial software [5,6] has supported the stable operation of the software for decades. However, there are still some technical challenges that are difficult to solve under the existing software architecture. The solution to these challenges requires the application of contemporary differential geometry and topology theories, which may not be compatible with the existing software architecture. Therefore, we propose to design a brand-new next-generation software architecture with super structured quadrilateral mesh generation as the core, so that the new generation of CAD/CAE industrial software can be more efficient.

## 3 Conclustion

Building on the application of contemporary frontier differential geometry and topology theories and algorithm design, this paper introduces the concept, research direction, and research field of super structured quadrilateral meshes with controllable overall arrangement structures of quadrilaterals. It presents preliminary illustrations of mesh generation research results and further poses a scientific question regarding the impact of super structured quadrilateral meshes on finite element computations. Investigating this scientific question will contribute to resolving the core technical challenges faced by current industrial software such as CAD/CAE and promoting their further development. Additional experimental data on super structured quadrilateral mesh generation (as shown in Figure 9) can be obtained via email [31].

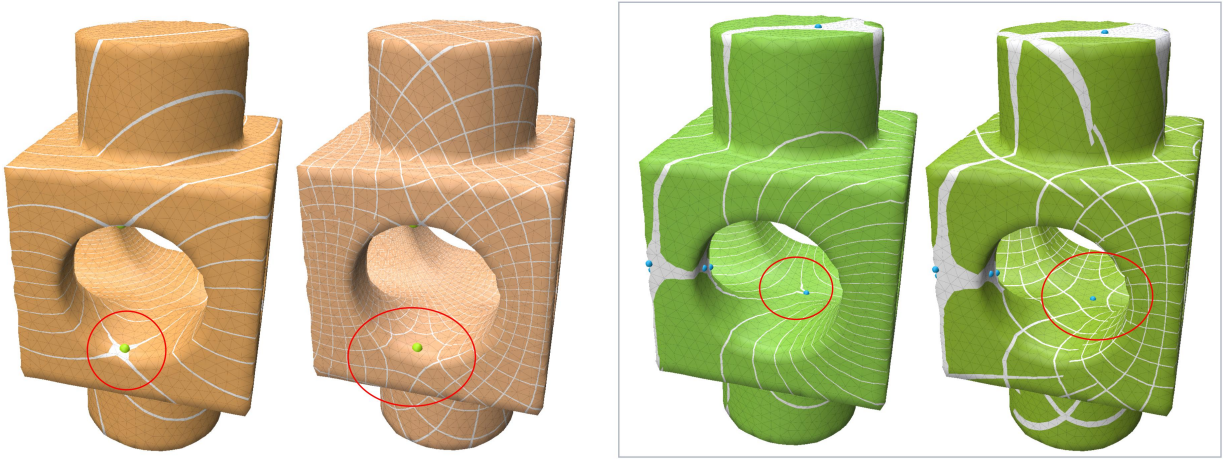

(Fig. 9: Structured vs Unstructured mesh, Generated by Hui Zhao with the Software Geometric.)

# References

1  Nico Pietroni, Marcel Campen, Alla Sheffer, Gianmarco Cherchi, David Bommes, Xifeng Gao, Riccardo Scateni, Franck Ledoux, Jean Remacle, and Marco Livesu. 2023. Hex-Mesh Generation and Processing: A Survey. ACM Trans. Graph. 42, 2 (April 2023), 1 – 44.

2  Hui Zhao, Shaodong Wang, and Wencheng Wang. 2022. Global Conformal Parameterization via an Implementation of Holomorphic Quadratic Differentials. IEEE Trans Vis Comput Graph 28, 3 (March 2022), 1529 – 1544.

3  Xianfeng Gu and Shing-Tung Yau. Global conformal surface parameterization。Proceedings of the 2003 Eurographics/ACM SIGGRAPH symposium on Geometry processing.

4  Teseo Schneider, Yixin Hu, Xifeng Gao, Jérémie Dumas, Denis Zorin, and Daniele Panozzo. 2022. A Large-Scale Comparison of Tetrahedral and Hexahedral Elements for Solving Elliptic PDEs with the Finite Element Method. ACM Trans. Graph. 41, 3 (June 2022), 1 – 14.

5  Altair.2022.HyperMesh. https://www.altair.com/hypermesh/

6  ANSYS.2022.ANSYS. https://www.ansys.com/products/meshing

7  Ryan Capouellez and Denis Zorin. 2024. Seamless Parametrization in Penner Coordinates. ACM Trans. Graph. 43, 4 (July 2024), 1 – 13.

8  David Bommes, Marcel Campen, Hans-Christian Ebke, Pierre Alliez, and Leif Kobbelt. 2013a. Integer-grid maps for reliable quad meshing. ACM Transactions on Graphics (TOG) 32, 4 (2013), 1 – 12.

9  David Bommes, Bruno Lévy, Nico Pietroni, Enrico Puppo, Claudio Silva, Marco Tarini, and Denis Zorin. 2013b. Quad-mesh generation and processing: A survey. In Computer graphics forum, Vol. 32. Wiley Online Library, 51 – 76.

10  David Bommes,HenrikZimmer,andLeifKobbelt.2009. Mixed-integer quadrangulation. ACMtransactions on graphics (TOG) 28, 3 (2009), 1 – 10.

11  Alon Bright, Edward Chien, and Ofir Weber. 2017. Harmonic Global Parametrization with Rational Holonomy.ACM Trans.Graph.36, 4 (2017).

12  Marcel Campen. 2017. Partitioning surfaces into quadrilateral patches: A survey. In Computer graphics forum, Vol. 36. Wiley Online Library, 567 – 588.

13  Marcel Campen, David Bommes, and Leif Kobbelt. 2015. Quantized global parametrization. Acm Transactions On Graphics (tog) 34, 6 (2015), 1 – 12.

14  Marcel Campen, Ryan Capouellez, Hanxiao Shen, Leyi Zhu, Daniele Panozzo, and Denis Zorin. 2021. Efficient and Robust Discrete Conformal Equivalence with Boundary. ACM Trans. Graph. 40, 6, Article 261 (dec 2021), 16 pages.

15  Marcel Campen and Denis Zorin. 2017. Similarity maps and field-guided T-splines: a perfect couple. ACM Transactions on Graphics (TOG) 36, 4 (2017), 1 – 16.

16  Ryan Capouellez and Denis Zorin. 2023. Metric Optimization in Penner Coordinates. ACMTrans. Graph. 42, 6, Article 234 (dec 2023), 19 pages.

17  Ryan Capouellez and Denis Zorin. 2024. Seamless Parametrization in Penner Coordinates. ACM Trans. Graph. 43, 4, Article 61 (jul 2024), 13 pages.

18  Hans-Christian Ebke, David Bommes, Marcel Campen, and Leif Kobbelt. 2013. QEx: robust quad mesh extraction. ACM Trans. Graph. 32, 6, Article 168 (Nov. 2013), 10 pages.

19  Mark Gillespie, Boris Springborn, and Keenan Crane. 2021. Discrete conformal equivalence of polyhedral surfaces. ACM Transactions on Graphics (TOG) 40, 4 (2021), 1 – 20.

20  Xianfeng Gu, Ren Guo, Feng Luo, Jian Sun, and Tianqi Wu. 2018a. A discrete uniformization theorem for polyhedral surfaces II. Journal of Differential Geometry 109, 3 (2018), 431 – 466.

21  Xianfeng Gu, Feng Luo, Jian Sun, and Tianqi Wu. 2018b. A discrete uniformization theorem for polyhedral surfaces. Journal of Differential Geometry 109, 2 (2018), 223 – 256.

22  Yixin Hu, Teseo Schneider, Bolun Wang, Denis Zorin, and Daniele Panozzo. 2020. Fast Tetrahedral Meshing in the Wild. ACM Trans. Graph. 39, 4, Article 117 (July 2020), 18 pages.

23  Yixin Hu, Qingnan Zhou, Xifeng Gao, Alec Jacobson, Denis Zorin, and Daniele Panozzo. 2018. Tetrahedral meshing in the wild. ACM Trans. Graph. 37, 4 (2018), 60 – 1.

24  Jingwei Huang, Yichao Zhou, Matthias Niessner, Jonathan Richard Shewchuk, and Leonidas J Guibas. 2018. Quadriflow: A scalable and robust method for quadrangulation. In Computer Graphics Forum, Vol. 37. Wiley Online Library, 147 – 160.

[25] Liliya Kharevych, Boris Springborn, and Peter Schröder. 2006. Discrete conformal mappings via circle patterns. ACM Trans. Graph. 25 (April 2006), 412–438. Issue 2.

[26] Zohar Levi. 2022. Seamless parametrization of spheres with controlled singularities. In Computer Graphics Forum, Vol. 41. Wiley Online Library, 57–68.

Zohar Levi. 2023. Seamless Parametrization with Cone and Partial Loop Control. ACM Transactions on Graphics 42, 5 (2023), 1–22.

[27] Hui Zhao and Steven J. Gortler. 2016. A Report on Shape Deformation with a Stretching and Bending Energy. arXiv:1603.06821 [cs] (March 2016).

[28] Hui Zhao, Kehua Su, Chenchen Li, Boyu Zhang, Lei Yang, Na Lei, Xiaoling Wang, Steven J. Gortler, and Xianfeng Gu. 2020. Mesh Parametrization Driven by Unit Normal Flow. Computer Graphics Forum 39, 1 (February 2020), 34–49. https://doi.org/10.1111/cgf.13660

[29] Hui Zhao, Xuan Li, Wencheng Wang, Xiaoling Wang, Shaodong Wang, Na Lei, and Xiangfeng Gu. 2019. Polycube Shape Space. Computer Graphics Forum 38, 7 (October 2019), 311–322. https://doi.org/10.1111/cgf.13839

[30] Hui Zhao, Na Lei, Xuan Li, Peng Zeng, Ke Xu, and Xianfeng Gu. 2017. Robust Edge-Preserved Surface Mesh Polycube Deformation. Pacific Graphics Short Papers (2017), 6 pages. https://doi.org/10.2312/PG.20171319

[31] Hui Zhao. WeChat Subscriber：Computational Discrete Global Geometric Structures. https://mp.weixin.qq.com/s/bydA_SMV-Hu1VHYOhrvsGA?scene=1.